\documentclass[apjl]{emulateapj}
 \slugcomment{Accepted to ApJL}
\usepackage{apjfonts}
 
\begin{document}

\title{A search for Lyman Break Galaxies at  \lowercase{$z >8$} in the NICMOS Parallel Imaging Survey
\footnotemark[1]}
 
 \author{Alaina L. Henry\altaffilmark{2},  Matthew A. Malkan\altaffilmark{2}, James W. Colbert\altaffilmark{3}, 
Brian Siana\altaffilmark{3}, Harry I. Teplitz\altaffilmark{3},  Patrick McCarthy\altaffilmark{4}, \& Lin Yan\altaffilmark{3}}
 
 \footnotetext[1]{Based on observations made with the NASA/ESA {\it Hubble Space Telescope}, obtained from the Space Telescope Science Institute, which is operated by the Association of Universities for Research in Astronomy Inc., under NASA contract NAS 5-26555.  These observations are associated with proposals 9484, 9865, and  10226.}

 \altaffiltext{2}{Department of Physics and Astronomy, Box 951547, UCLA, Los Angeles, CA 90095, USA; ahenry@astro.ucla.edu, malkan@astro.ucla.edu                       }
 \altaffiltext{3}{$Spitzer$ Science Center, California Institute of Technology, 220-6, Pasadena, CA, 91125, USA; colbert@ipac.caltech.edu, bsiana@ipac.caltech.edu, hit@ipac.caltech.edu, lyan@ipac.caltech.edu}
 \altaffiltext{4}{Observatories of the Carnegie Institute of Washington, Santa Barbara Street, Pasadena, CA 91101; pmc2@ociw.edu}

 \begin{abstract}
We have selected 14 J-dropout Lyman Break Galaxy (LBG) candidates with $J_{110} - H_{160} \ge 2.5$ from the NICMOS Parallel Imaging Survey.
 This survey consists of 135 square arcminutes of imaging in 228 independent sight lines, reaching average $5\sigma$ sensitivities of $J_{110} = 25.8$ and $H_{160} = 25.6$ (AB).
 Distinguishing these candidates from dust reddened star forming galaxies at $z\sim 2-3$ is difficult, and will require longer wavelength observations.    We consider the likelihood that any J-dropout LBGs exist in this survey, and find that if $L^*_{z=9.5}$ is significantly brighter than $L^*_{z=6}$ (a factor of four), then a few J-dropout LBGs are likely.  A similar increase in luminosity has been suggested  by Eyles et al. and Yan et al., but the magnitude of this increase is uncertain.      \end{abstract}

 \section{Introduction}
Searches for Lyman break galaxies (LBGs) have measured their luminosity
functions out to high redshifts ($z=4-5$; \citealt{Kashikawa}; \citealt{Yoshida}).  This indicates that substantial numbers of
galaxies were in place at z=6, and that galaxy formation must be ongoing at higher redshifts.  \cite{Yan06} and \cite{Eyles} have shown that the existing mass in $i$-dropouts at $z=6$ can not be assembled unless star formation rates were higher in the past.   The implied rapid first bursts of star formation at $z>7$ could produce galaxies with large enough rest-wavelength UV luminosities to be detected  in the near infrared.  In addition, it is currently uncertain if the the amount of star formation at $z=6$ is adequate to reionize the universe.  J-dropout LBGs  may play an important role in this process,  which is likely between  $9 \le z \le 14$ \citep{Spergel}.

The NICMOS Parallel
    Imaging Survey (\citealt{nicparposter}; Henry et al. in prep), 
    provides the unique combination of
  sensitivity and wide area coverage required to find rare, luminous  LBGs with $8 \le z \le 11$.         
    With 135 arcminutes$^2$
      in 228 independent sight lines, we cover an order of magnitude more 
      sky than the NICMOS portions of  Hubble Deep Field North
       (HDF-N;  \citealt{Dickinson99}; \citealt{Thompson99}) and the Ultra Deep Field 
       (UDF; \citealt{Thompson05}).  Although the NICMOS Parallel Imaging Survey is less sensitive than these surveys, the galaxies presented here are better suited for follow-up observations.      In addition, it surpasses most ground based imaging.  
 
 We have identified a sample of possible J-dropout galaxies
  from the NICMOS Parallel Survey, and here we discuss the likelihood that 
  any of them are at $z \ge 8$, as well as the possible implications for star
  formation near the epoch of reionization.     We use 
  $H_o = 71 ~{\rm km~ s}^{-1} ~{\rm Mpc}^{-1}$, $\Omega_{\Lambda} = 0.73$, and $\Omega_M = 0.27$, and AB magnitudes are used throughout.   We take $M^*_{1500}(z=3) = -21.1$ \citep{Steidel99}, which corresponds to $H_{160} = 26.4$ at $z=9.5$.

\begin{figure*}
\plottwo{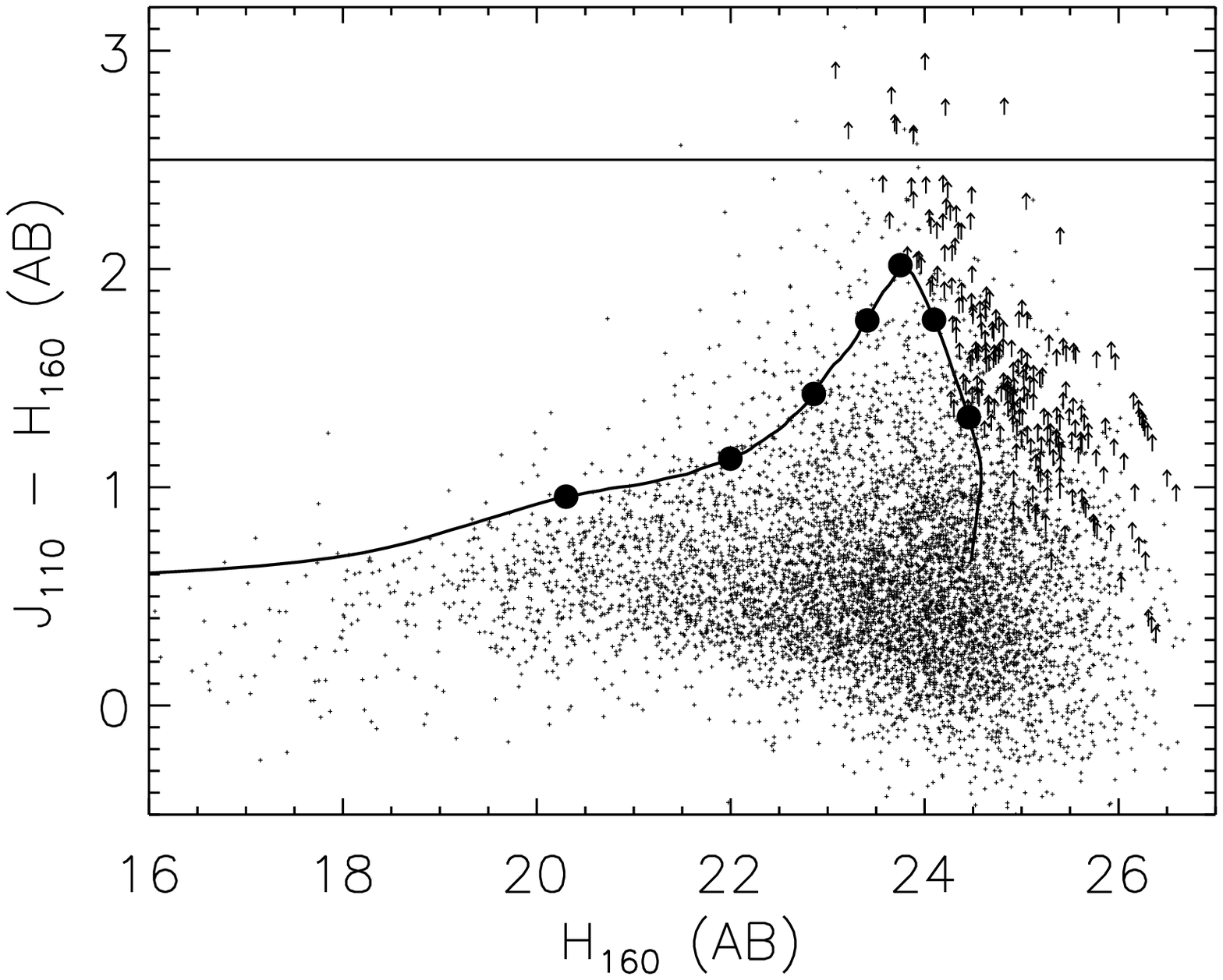}{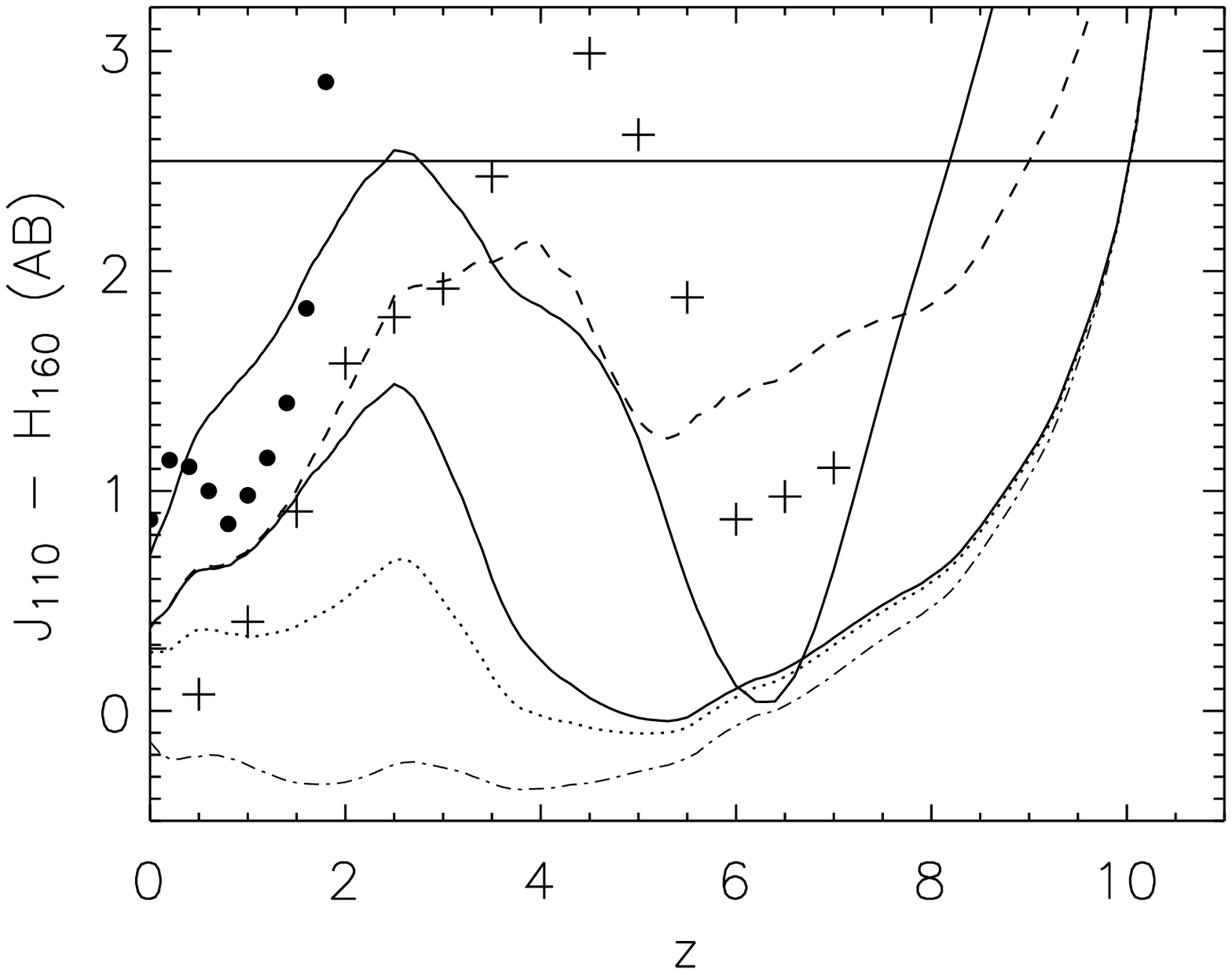}
\caption{ {\it Left --} The NICMOS Parallels Imaging Survey color magnitude diagram.    All points are  $5\sigma$ detections, or better, in $H_{160}$.  Arrows represent sources with
  no $J_{110}$ detection above $3\sigma$. 
We overplot the \cite{bc03} evolutionary track of a galaxy which formed at $z_f=12$, with 
a $A_V = 1$, a Salpeter IMF, solar metallicty, and an exponentially declining star formation rate with a 500 Myr e-folding time.  This model galaxy has a stellar mass of $7 \times 10^{10} M_{\sun}$  at $z=0$.  
The track is marked with fiducial redshift points at $z=0.5, 1.0, 1.5, 2.0, 2.5~ {\rm(peak)}, 3.0, 3.5.$ 
    {\it Right} --Curves show \cite{bc03}
   evolutionary tracks.  All models have a Salpeter initial mass function and solar metallicity.  The bluest, dot-dashed track is a 10 Myr old model with constant star formation and no extinction, while the remaining tracks are evolving, with $z_f = 12$.    Their star formation histories and extinctions are: {\it dotted--} Constant with 
  $10~ M_{\sun} ~{\rm yr}^{-1}$, and A$_V =0$; {\it solid--} Exponentially declining, with a 0.5 Gyr e-folding time,
   and A$_V =0$ and 2; {\it dashed--} An instantaneous burst with A$_V=0$.   Extinction is added following the prescription by \cite{Cardelli}, which contains a 2000\AA\ bump.  At $z\sim6$ this feature is in the $H_{160}$ filter, the resulting dust obscured color is {\it bluer} than the intrinsic color.
   We also show the expected colors from two local ULIRG SEDs, Arp 220 (crosses), and IRAS 03521+0028 (points).    
   The horizontal line indicates the color cut which we made at $J_{110} - H_{160} \ge 2.5$.  
}  
\label{cmd}
\end{figure*}

 \section{Observations \& Source Selection}
 \subsection{The NICMOS Parallel Survey}
 We conducted the NICMOS Parallel Survey in  {\it Hubble Space Telescope} (HST) observing cycles 7, 11, 12, and 13.   Images and slitless grism spectroscopy were obtained in the parallel mode, using the NICMOS 3 camera \citep{Thompson98}.   Early results from the cycle 7 survey are presented in \cite{McCarthy},  Yan et al. (1998, 1999, 2000),  and \cite{Teplitz}.   
 
 The cycles 11, 12, and 13 imaging data will be described in more detail in Henry et al. (in prep);
  we provide a summary here.   
  Each field was observed with both
the $J_{110}$ ($F110W$) and $H_{160}$ ($F160W$)  filters, with average integration times
 of $\sim$1500~s  and $\sim$2000~s.  Images are drizzled according to \cite{drizzle}, and the resulting
 point spread function has a FWHM $\sim0.3\arcsec$.  Photometry
  was performed with SExtractor \citep{sextractor} in dual image
   mode.  The $H_{160}$ images were used to detect sources, and fluxes were measured from identical \cite{kron} apertures in both bands.  We require 
   9 pixels above 1.5$\sigma$ for detection.    We further
    require a 5$\sigma$ detection in $H_{160}$, as measured in 
    a 0\arcsec.8 diameter aperture.    The average 5$\sigma$ sensitivities 
       reached are J$_{110}$ = 25.8 and H$_{160}$ = 25.6.   We detect 7700 objects; the  color magnitude diagram is shown in  Figure \ref{cmd} (left).

    \subsection{J-dropout candidates}
  \label{sample}
 In Figure \ref{cmd} (right) we show \cite{bc03} evolutionary tracks to which we have added an IGM foreground absorption component which is completely opaque beyond $z=6$,  where Ly$\alpha$ passes into the $J_{110}$ filter.
   In addition, we show examples of two local ultra-luminous infrared galaxies (ULIRGs), Arp 220, and IRAS 03521+0028.  For Arp 220 we compile the UV to near infrared spectral energy distribution (SED) from \cite{Goldader}, \cite{Surace00}, and the RC3 catalog.  For IRAS 03521+0028, which has a very red rest frame optical SED, only the BIHK' photometry from \cite{Surace00} is available, so we do not extend this track beyond $z=2$.

To minimize contamination from lower redshift interlopers, we adopt the color selection $J_{110} - H_{160} > 2.5$ for J-dropout candidates.  Figure \ref{cmd} (right) shows that the reddest J-dropout LBGs will be selected at $z\ga 8$, while nearly all models meet this criterion at $z> 10$.   We find  14 J-dropout candidates (Table \ref{table}); five are detected in $J_{110}$ (at $3\sigma$ or better), but this does not disqualify them as LBGs, since some flux will fall in the $J_{110}$ filter for $z < 10.5$.   The remaining nine galaxies are among the sources which are plotted as lower limits (upward pointing arrows) in Figure \ref{cmd}.   
Possible contamination from lower redshift sources is discussed in \S \ref{interlopers}.

 Since exposure times varied from field to field,  the NICMOS Parallel Survey is inhomogeneous in sensitivity.  For selection of red galaxies, the faint limit of the sample is determined by the sensitivity in the $J_{110}$ images.  For example,   at  
 $H_{160}=23.3$, 98\% of the $J_{110}$ images have sufficient sensitivity for sources to be selected with $J_{110} - H_{160} \ge 2.5$. A half magnitude fainter, at $H_{160}=23.8$, this completeness drops to $\sim 50$\%.  Since fainter galaxies are more likely to be genuine J-dropout LBGs, we consider all galaxies with $J_{110}-H_{160} \ge 2.5$, and correct  for this incompleteness.

\section{Discussion} 
 \subsection{Foreground Contamination}
\label{interlopers}
Figure \ref{cmd} (right) shows that some lower redshift galaxies may be selected as J-dropout candidates.  
\cite{Bouwens05} find three galaxies with $J_{110} - H_{160} \ge 2.5$ in a 15 arcmin$^{2}$ survey.  
This density is consistent with the 14 sources we have found, since the NICMOS Parallel Imaging Survey covers ten times the area of Bouwens et al., although with less sensitivity. 
Bouwens et al. use optical and K-band photometry to argue that these galaxies are all at lower redshift.  Since they find no J-dropout LBGs with  $J_{110} - H_{160} \ge 2.5$, the implied density is  less than 250 degree$^{-2}$.

In addition, we consider the SEDs of two local ULIRGs.  \cite{Daddi} have shown that ULIRGs have a surface density of about $1~{\rm arcmin}^{-2}$ at $z=2-3$.  For  Arp 220 ($L_{bol} = 1.5\times 10^{12} L_{\sun}$), we expect $H_{160} = 23.6$ at $z=2$, which is typical of the galaxies in this survey.  However,  ULIRGs like Arp 220 are not red enough at $z=2-3$ to be selected as part of this sample.  
In Figure \ref{cmd} (right) we show the predicted color of another ULIRG, IRAS 03521+0028, which has a very red rest frame optical SED. This galaxy is red enough to be selected at $z=2$, but it is doubtful if many ULIRGs at $z=2-3$ can be described by this red optical SED.  If $\sim$10\%  display similar colors, this could account for the 14 J-dropout candidates.  While galaxies like Arp 220 could also be selected by our color cut at $z\sim4-5$,  ULIRGS at these redshifts are expected to be rare, and would have to be very luminous, with $L_{bol} \ga \times 10^{13} L_{\sun}$.  Even at $z=2-3$, the density of ULIRGs of this luminosity is low, with only $\sim 15~ {\rm degree}^{-2}$  \citep{LYan05}.  

  Contamination from Galactic sources is improbable.    \cite{Dickinson00} report on a  similar point source in 
the HDF-N, with $J_{110} - H_{160} > 2.3$, and find that it is unlikely a stellar or substellar source.

\begin{figure*}
\epsscale{1}
\plotone{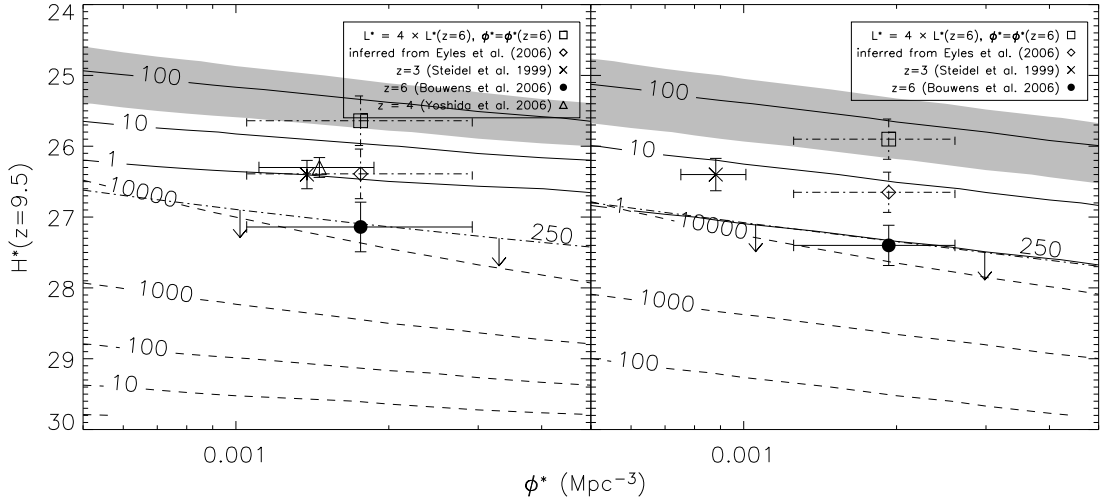}
\caption{Contours of galaxies per square degree 
for this survey (solid) and a survey which is complete to  28.0 (dashed), calculated
 using a Schecter luminosity function (left), and a broken power law with the high-L slope fixed at -4 (right).      The grey region indicates 1-10 objects in 
 this 135 square arcminute survey.  The dot-dashed curve indicates $\la 250~ {\rm degree}^{-2}$ in the \cite{Bouwens05} search for J-dropout LBGs.  In the right panel, the dot-dashed curve overlaps the solid contour for $1~ {\rm degree}^{-2}$.      The points indicate estimates of the LF at z=9.5, from lower redshift populations.  }
\label{lfcontour}
\end{figure*}

 \subsection{Estimating the Density of Galaxies at $z\ge8$}
 \label{density}
Since no objects have ever been confirmed at less than 
500 Myrs after the Big Bang, it is difficult to predict how bright
 or numerous they will be.  Since small uncertainties in $L^*_{z=9.5}$ can have a large effect,  we consider both a Schecter luminosity function (LF), and a 
 broken power law of the form:  
\begin{equation}
{\phi (m) } =
{\phi^*  \over 10^{0.4 ( \alpha_1 +1) ( m-m^*)} + 
10^{0.4 (\alpha_2 +1) (m-m^*)} }
\end{equation}
with faint end slope $\alpha_1$ and luminous slope $\alpha_2$\footnotemark[1].  
\footnotetext[1]{
 We fit the above broken power law to the $z=3$ population from \cite{Steidel99}, 
 and the $z=6$ LBGs from \cite{Bouwens06}. We find $\alpha_2 \sim -4$, and $\alpha_1 \sim -1.6$  in both populations, so we hold both slopes constant.   
}
In Figure \ref{lfcontour}, we show contours in surface density, calculated by integrating the luminosity function for a range of plausible $\phi^*$ and $H^*(z=9.5)$: 
 \begin{equation}
 N = \int_z dz {dV \over dz} \int_m P(m, z) \phi(m, m^*(z)) dm .  
 \end{equation}
 This allows for the dimming of $H^*_{160}$ with redshift, which is large for $z>10.5$, where the Ly$\alpha$ absorption passes into the $H_{160}$ filter.
The quantity $dz {dV \over dz}$ is the comoving volume element, and $P(m, z)$ corrects for galaxies which are missed because of incompleteness, or because they are not red enough to be selected by our color cut.  To estimate $P(m, z)$, we separate the quantity into $P(m, z) = p_1(m) p_2(z)$, taking $p_1(m)$ as the completeness which we discussed in \S \ref{sample}.  For $p_2(z)$, we assume that galaxies are equally distributed between the reddest and bluest models in Figure \ref{cmd} (right)\footnotemark[2], so that  $p_2(z)$ increases from zero to one between $8 < z < 10$.
\footnotetext[2]{This is consistent with the results of \cite{Thompson03} and \cite{Thompson06}, which find that high redshift galaxies are best fit by the bluest spectral types.  Even the reddest model used here is only a few hundred Myrs old at $z=8$ and is therefore comparable to the blue models used by Thompson et al.     
}
    Under these assumptions, the galaxies in this survey should lie between $8 \le z \le 11$, with an average redshift of  between $z=9-10$, depending on the chosen LF.  We therefore adopt an average redshift  of $z=9.5$.  We repeat this calculation to estimate a constraint on the LF based on the upper limit in surface density found in the \cite{Bouwens05} search for J-dropouts, which consists of six fields imaged to an average sensitivity of $H_{160} = 28.0$.  Like the NICMOS Parallel Survey, the \cite{Bouwens05} survey is limited by sensitivity in $J_{110}$, so that galaxies with $J_{110} - H_{160} \ge 2.5$ are only selected down to $H_{160} \sim 25.5$.  Their finding of less than 250 degree$^{-2}$ is plotted as the dot-dashed line in Figure \ref{lfcontour}.   Also, for comparison to more sensitive surveys, we show surface densities calculated with $p_1(m) =1$ and a detection limit of $H_{160} = 28$ (This requires a sensitivity of $J_{110} = 30.5$, which can be reached in less than 10 hours of integration with the {\it James Webb Space Telescope}).

 These plots show 
 the expected density of J-dropout LBGs,  based on 
 a given $H^*$ and $\phi^*$.  
 The shaded area of Figure \ref{lfcontour} shows the allowed region for 
 $H^*$ and $\phi^*$ if 1-10 J-dropout LBGs exist in this survey. 
  If none of the candidates
 are $z\ge 8$ LBGs, then $H^*$ and $\phi^*$ should be 
 below the shaded region.    
  We show examples from a few observed 
 LFs, including the broken power law fits which we made.  If the $z=6$ measurement by \cite{Bouwens06} 
is a good indicator of  the population at $z\sim9.5$, then the density of sources which we expect is of order $0.01 ~{\rm degree}^{-2}$ in the Schecter form, and about $1 ~{\rm degree}^{-2}$ in the broken power law LF.  
These densities are too low to suggest any J-dropout LBGs in the NICMOS Parallel Survey, which covers only 0.04 ${\rm degree}^{2}$.      
On the other hand, \cite{richard}, in a lensing survey for J-dropout galaxies, find that the $z=8-10$ LF may be more like the $z=3$ LF by \cite{Steidel99}.  In this case, the chances of finding a J-dropout LBG in the NICMOS Parallel Survey are better, with about 1-10 ${\rm degree}^{-2}$ in both the Schecter and broken power law forms.    

The LF constraint which we estimated from the absence of J-dropout galaxies with $J_{110} - H_{160} \ge 2.5$ in the \cite{Bouwens05} survey suggests they are too few and too faint for any to be found in the NICMOS Parallel Survey.   However, Bouwens et al.  survey only a small volume for rare objects, and so the cosmic variance will be large.   Alternatively,  the results of  \cite{Yan06} and \cite{Eyles}  show that  to produce the observed mass density at $z=6$, the progenitors at $z>7$ must have formed stars more rapidly in the past.   The magnitude of this brightening is uncertain, but \cite{Eyles} suggest it is a factor of  a few.  We show this inferred LF, as two times brighter than the $z=6$ measurement by \cite{Bouwens06}, with no density evolution.  This predicted LF is similar to the measurement by \cite{richard},  again, suggesting that we expect, at best, a 40\% chance of finding a J-dropout LBG.    If the brightening in the LF is instead a factor of four larger than at $z=6$, then we can expect one or two J-dropout LBGs.     In addition,  we find that more than a factor of ten increase in $L^*$ (with no density evolution) is unlikely, since we would then expect more than 16 galaxies with $J_{110} - H_{160} \ge 2.5$ in this survey.

\subsection{Star Formation   \& Reionization at $z\ge8$}

J-dropout galaxies may play an important role in the reionization of the universe.
The critical star formation rate density ($\rho_{SFR}$) required to ionize the universe,  is  $\rho_{SFR}(z=9.5)=0.09~ M_{\sun}~{\rm yr}^{-1}~{\rm Mpc}^{-3}$ \citep{MHR99}, if all ionizing photons escape the galaxies, and the clumping factor of ionized hydrogen is 30.   
These parameters are uncertain, 
but this is much larger than the upper limit posed by \cite{Bouwens05} for galaxies at $z\sim 10$.   Alternatively, for the LF which predicts  a few J-dropout LBGs in this survey, where $L^*_{z=9.5} = 4 \times L^*_{z=6}$, and $\phi^*_{z=9.5} = \phi^*_{z=6}$, we calculate $\rho_{SFR}(z=9.5) \sim 0.1~ M_{\sun}~{\rm yr}^{-1}~{\rm Mpc}^{-3}$ in both the Schecter and broken power law forms.   In this case, J-dropout LBGs may be capable of reionizing the universe, although the uncertainties in the escape fraction of ionizing photons, and the clumping factor of ionized hydrogen must be addressed.  

To conclude, it is not implausible that $\rho_{SFR}$ is very large during the first few hundred million years of galaxy formation if stars were formed in rapid bursts.  In fact, cofirmation of any J-dropout galaxies will suggest a large $\rho_{SFR}$ for most reasonable  $H^*(z=9.5)$.  Follow-up observations of the galaxies presented here is one way to constrain star formation at $z\ge8$.

\acknowledgments
 This research was supported by NASA through {\it Hubble Space Telescope} Guest Observer grants 9865 and 10226.  
 We are grateful to R. Thompson  for helpful comments which improved this manuscript.

\begin{deluxetable}{ccccccc}
\tablecolumns{7}
\tablecaption{J-dropout LBG candidates with $J_{110} - H_{160} > 2.5$. }
\tablehead{ 
\colhead{ ID} & \colhead{ $H_{160}$} & \colhead{$H_{160}$ S/N }& \colhead{$J_{110} - H_{160}$  (AB)}  &  \colhead{RA (J2000)}  & \colhead{Dec(J2000)} &\colhead{ FWHM (\arcsec)}
}
\startdata
1  &    24.8    &   36     &   $>2.7$     & 14 46 16.09     & 40 30 10.9   &  0.32 \\  
2  &    23.2    &   26     &  $>2.6$      & 22 15 11.95     & -14 08 01.0  & 0.55\\
3  &    23.9    &    26    &  $>2.6$  &   02 05 29.41    &  -12 05 55.1  & 0.39 \\
4  &   24.2    &     35	& $ >2.7$	& 02 31 17.39     & -08 49 17.5  &  0.30 \\
5  &   23.3     &  50    &  $ ~~~3.2$     &   00 59 36.34    & -51 13 55.5   & 0.44\\
6  &  23.9   &   30    &   $> 2.6$   &  07 44 21.68     & 39 21 41.1   & 0.42\\
7  &   23.7   &  37    & $ > 2.8 $  &  09 48 39.80   & 67 30 23.7   &   0.32 \\
8 & 24.0   &    43      &  $> 2.9$    & 10 47 44.77  & 13 46 58.1    &  0.46 \\
9  & 23.9  &      28   &  $~~~2.6$    &     15 34 13.04  & 26 48 24.3 &  0.36\\
10  & 23.2  &    48    &   $~~~3.1$    & 12  35 42.37  & 62 10 30.4   &  0.38 \\
11   &  22.7  &  51    &  $~~~2.7$    &  02 09 57.31  & -50 55 29.2   &  1.47 \\
12  &  23.7  & 27  &  $>2.6$  & 21 03 57.60  &  03 16 28.2 & 0.37 \\
13  &  23.8   &30       & $~~~2.6$  & 17 58 07.12  & 66 45 19.9 & 0.49 \\
14  & 23.1  &  33 &     $ > 2.8$  &  00 25 42.99 & -12 25 29.2  & 0.39 \\
\enddata
\label{table}
\end{deluxetable}

\end{document}